\documentclass[aps,prd,twocolumn,superscriptaddress,english]{revtex4-1}
\usepackage[T1]{fontenc}
\usepackage[utf8]{inputenc}
\usepackage{amsmath}
\usepackage{amssymb}
\usepackage{graphicx}
\usepackage{bm}
\usepackage{bbm}
\usepackage{subfigure}
\usepackage{float}
\usepackage{amsfonts}
\usepackage{color}
\usepackage{txfonts}
\usepackage{array}
\usepackage[english]{babel}
\usepackage{soul}

\makeatletter
\usepackage[colorlinks = true,
            linkcolor =red,
            urlcolor  =magenta,
            citecolor = blue,
            anchorcolor = blue]{hyperref}

\makeatother
\usepackage{babel}
\begin{document}
\title{Vacuum pair production under spatially asymmetric time-oscillating electric fields}
\author{Mamat Ali Bake}
\email[Electronic mail:]{mabake@xju.edu.cn}
\affiliation{Xinjiang Key Laboratory of Solid State Physics and Devices, School of Physics Science and Technology, Xinjiang University, Urumqi 830017, China}
\author{Obulkasim Olugh}
\affiliation{Department of forensic science and technology, Xinjiang Police College, 830011, Urumqi, China}
	
\begin{abstract}
We investigate electron-positron pair production from the quantum vacuum in spatially asymmetric, time-oscillating electric fields using the Dirac-Heisenberg-Wigner (DHW) formalism. The field configuration combines spatially separated Sauter-type pulses with temporal oscillations, including frequency chirps and phase modulation. Our results demonstrate that spatial asymmetry significantly enhances pair production compared to symmetric fields, while optimal tuning of temporal parameters (e.g., frequency $\omega$ and chirp $b$) further amplifies the yield. For $\omega \gtrsim 0.4m$, multiphoton-dominated processes generate oscillatory momentum spectra, whereas low-frequency fields ($\omega \lesssim 0.3m$) exhibit tunneling-dominated Gaussian distributions. Chirped fields induce spectral asymmetry and interference patterns, with peak yields increasing by up to a factor of 9 for $\omega = 0.7m$ and $b = 0.5\omega/\tau$. These findings provide a pathway to optimize pair production in experimentally feasible spatiotemporal field configurations.
\end{abstract}
\maketitle
	
\section{INTRODUCTION}

The vacuum electron-positron pair production is one of the most fascinating prediction of quantum electrodynamics (QED), a theory that describes how light and matter interact at a fundamental level. In the framework of QED, the vacuum is not a static void but rather a dynamic medium characterized by the presence of virtual particle-antiparticle pairs, which continuously undergo processes of generation and annihilation as a result of quantum fluctuations. When an extremely strong electromagnetic field is present, it can provide enough energy to convert these virtual particles into real, observable particles, namely, electrons and positrons.

The possibility of pair creation was first pointed out by Sauter in 1931 \cite{Sauter:1931zz} as a resolution to the Klein paradox in relativistic quantum mechanics \citep{Klein}. Sauter`s idea was extended to QED by Schwinger in 1951 \citep{Schwinger:1951nm} and explicitly computed the vacuum persistence probability using the proper time method in a constant electric field in terms of the one-loop Heisenberg-Euler effective action. Then, the number of created pairs was first computed by Nikishov \citep{Nikishov} by solving the Dirac equation in a constant electric field and using the Green function technique \citep{Review2023}.
However, this effect has not yet been observed experimentally due to the extremely high electric field strength $E_{cr}\approx 1.3\times 10^{18}~\textrm{V}/\textrm{m}$ (the so-called Schwinger critical field strength; the corresponding laser intensity is $I_{cr}\sim 4.3 \times 10^{29}~\text{W/cm}^{2}$) required to produce electron-positron pairs. With the rapid development of laser technology, the intensity of high-power lasers is expected to approach $I\sim 10^{26}~\text{W/cm}^{2}$ in the near future. This will allow researchers to observe the nonperturbative Schwinger effect in the laboratory in the near future \citep{eli-beams,xcels}.

There are two primary mechanisms for vacuum electron-positron pair production: tunneling \citep{tunnel} and the multiphoton process \citep{multiphoton}. The tunneling mechanism dominates in the presence of a strong, low-frequency electric field. This process is analogous to a particle overcoming an energy barrier that it classically should not be able to cross. Whereas the multiphoton effect becomes more significant in a high-frequency, weak-field scenario. In this case, multiple photons interact coherently to accumulate enough energy to create an electron-positron pair. Therefore, for a high-frequency fields, even if weaker in magnitude than the Schwinger critical field strength, can still facilitate pair production due to the cumulative effect of photon interactions. These two mechanisms are distinguished by the Keldysh parameter $\gamma = {m_e\omega} / {eE_0}$~\citep{Keldysh:1965ojf} (where $m_e$, $e$, $\omega$, and $E_0$ are the electron mass, charge, frequency, and external electric field strength, respectively). For $\gamma \ll 1$, pairs are primarily produced through the tunneling process, while for $\gamma \gg 1$, the pairs are mainly created via the multiphoton effect. Multiphoton effect was observed in SLAC experiments by the collision of a high-energy electron beam with an intense laser pulse \citep{SLAC1,SLAC2}. However, the tunneling process has not yet been experimentally explored because of the very high Schwinger critical field strength.

Another promising approach for pair production under subcritical field strengths is the dynamically assisted Schwinger mechanism \citep{Schutzhold:2008pz,Taya:2020,CK:2021}. This mechanism involves the combination of two electric fields, one characterized by high intensity and low frequency, and the other by high frequency and low intensity. The high-intensity field facilitates the tunneling effect, while the high-frequency field drives the multiphoton process, where the Keldysh parameter satisfies $\gamma \sim 1$. Although the tunneling and multiphoton effects are individually suppressed, their combined influence significantly enhances the pair production rate through dynamical assistance, effectively reducing the required field intensity \citep{Aleksandrov2018,IbrahimSitiwaldi2018}. Therefore, recent studies have focused on enhancing the pair production rate through various combinations of electric fields, leveraging the dynamically assisted Schwinger mechanism \citep{Ibrahim:2023,Olugh:2019nej,Aleksandrov2018,Linder:2015vta,Taya:2020,CK:2021,Copinger:2016,Torgrimsson:2019}.
Meanwhile, many recent studies have demonstrated that the pair production rate and particle dynamics are sensitive to the external field parameters, such as the pulse duration, carrier-envelope phase, pulse number, and spatial scale \citep{Hebenstreit:2011wk,Kohlfurst:2015niu,Ababekri:2019dkl,Aleksandrov:2016,Aleksandrov:2017,Aleksandrov:2020,Hebenstreit:2009km, Dumlu:2010vv,Dumlu:2010ua,Dumlu:2011rr,Olugh:2018seh,Kohlfurst:2015zxi,Mamutjan:PLB,Malika:2023,Emin:2023}. Therefore, optimizing the external field configuration is crucial in vacuum pair production studies.

Although many previous studies have confirmed that dynamically assisted mechanisms can reduce the required field intensity, most of them have primarily focused on spatially homogeneous time-dependent external fields rather than space-dependent fields. This is because the Schwinger effect is generally suppressed in spatially inhomogeneous electric fields \citep{Dunne:2005,Ilderton:2014,Ruf:2009,Gies:2016,Gies:2017}. Specifically, spatial inhomogeneities provide momentum to the Dirac sea, effectively increasing the mass gap and thereby suppressing pair production efficiency. In contrast, temporal inhomogeneities typically enhance the pair production effeciency by supplying energy to the Dirac sea \citep{Gelis:2016}.
Another effect of spatially inhomogeneous fields is the acceleration of pairs by the electric field immediately after their creation. Depending on the spatial extent of the electric field, the momentum acquired by the pairs upon creation varies. Consequently, the spatial inhomogeneities of the field influence the dynamics following pair creation, thereby affecting momentum spectra \citep{Emin:2023,Amat:2023vwv,Bake:2024} and leading to phenomena such as self-bunching \citep{Hebenstreit:2011wk,Ababekri:2019dkl} and ponderomotive effects \citep{Ck:2018}. Additionally, these spatial inhomogeneities also modify temporal effects, including the dynamically assisted Schwinger effect and quantum interference.

Notably, electric fields realized in real physical scenarios or experimental studies are inherently inhomogeneous in both space and time. These inhomogeneities can substantially influence the vacuum pair production process. Therefore, some recent studies have investigated space-dependent electric fields and attempted to mitigate the suppression of pair production efficiency from spatial inhomogeneity by employing a combination of spatially inhomogeneous, time varying electric fields \citep{Schneider:2016,Ababekri:2019dkl,Aleksandrov2018,Liliejuan:2021,Malika:2021,Emin:2023}.
However, the majority of studies have focused on symmetric spatially inhomogeneous electric fields due to the convenience of simplifying theoretical and numerical calculations. Additionally, it is well-known that, the temporal asymmetry of the external field has a significant impact on vacuum pair production \citep{Oluk:2014,Oluk:2020,Chen:2024}. Nevertheless, the effects of spatial asymmetry of the electric field on pair production remain poorly understood and warrant further investigation.

In this study, we conducted a systematic investigation into vacuum pair production in spatially and temporally inhomogeneous electric fields. Specifically, we examined a spatially asymmetric electric field with a temporal component characterized by various fundamental frequencies, frequency chirps, and initial phases. When employing a spatially asymmetric field configuration, the interplay between the spatial parameters of the field becomes more complex compared to the symmetric case, leading to distinct outcomes as the frequency and spatial scale of the field vary. Additionally, the results are sensitive to the initial phase of the field. Notably, we identified an optimal field combination mode that could be utilized in future strong-field QED experiments.

The remainder of this paper is structured as follows. In Section \ref{method}, we provide a concise review and present the key equations for the Dirac-Heisenberg-Wigner (DHW) formalism (Section \ref{DHWformalism}). Additionally, we introduce a model of a spatially asymmetric external field with temporal oscillation (Section \ref{fields}). In Section \ref{results}, the results of DHW calculations for spatially symmetric and asymmetric fields are compared first. Subsequently, comparisons of results obtained under varying spatial scales, frequencies, chirp parameters, and initial phases are presented, along with discussions of their physical implications. Finally, the main conclusions are summarized in Section \ref{summary}.

We used natural units throughout this paper, where $\hbar = c = 1 $, and all physical quantities were expressed in terms of the electron mass $m$. For example, the units of the field frequency and momentum are $m$, and the time scale of the electric field is $1/m$.

\section{method and external field description}\label{method}

\subsection{DHW formalism}\label{DHWformalism}
The DHW formalism, a relativistic phase-space approach, is widely used to study vacuum pair production. It efficiently investigates pair creation under time-dependent and spatially varying electromagnetic fields. In this framework, the external electromagnetic field is approximated by its mean field using a Hartree-type approximation, and particles are treated as quantum fields. For more details on calculations and numerical strategies, see Refs.~\citep{Hebenstreit:2011wk,Kohlfurst:2015zxi}. Below, we provide a brief introduction to the key concepts of the DHW method.

First, in the case where a background field exists, we write the Lagrangian in the QED
\begin{equation}\label{Lagrangian}
\begin{aligned}
L(\Psi ,\bar{\Psi },A)=\frac{1}{2}\left( \text{i}\bar{\Psi }{{\gamma }^{\mu }}{{D}_{\mu }}\Psi -\text{i}\bar{\Psi }D_{\mu }^{\dagger }{{\gamma }^{\mu }}\Psi  \right) -m\bar{\Psi }\Psi -\frac{1}{4}{{F}_{\mu \nu }}{{F}^{\mu \nu }}
\end{aligned},
\end{equation}
where ${{\mathcal{D}}_{\mu }}=\left( {{\partial }_{\mu }}+\text{i}e{{A}_{\mu }} \right)$ and $\mathcal{D}_{\mu }^{\dagger }=\left( \overleftarrow{{{\partial }_{\mu }}}-\text{i}e{{A}_{\mu }} \right)$ are the covariant derivatives with a vector potential ${{A}_{\mu }}$ that vanishes at asymptotic times, and ${{\gamma }^{\mu }}$ are the gamma matrices.
To describe the dynamics of particles, the Dirac equation is essential. By applying the Euler-Lagrange equation to the Lagrangian given in Eq.~\ref{Lagrangian}, both the Dirac equation and its adjoint equation can be expressed as follows:
\begin{equation}\label{DiracEq}
\begin{aligned}
\left( {{\rm{i}}{\gamma ^\mu }{\partial _\mu } - e{\gamma ^\mu }{A_\mu } - m} \right){\rm{\Psi }} = 0,\\
{\rm{\bar \Psi }}\left( {{\rm{i}}{\overleftarrow{{{\partial }_{\mu }}} }{\gamma ^\mu } + e{\gamma ^\mu }{A_\mu } + m} \right) = 0.
\end{aligned}
\end{equation}
Subsequently, by employing the Dirac spinor ${\rm{\Psi }}$ and ${\rm{\bar\Psi}}$, we constructed a gauge-covariant density operator within the Heisenberg picture:
\begin{equation}\label{DensityOperator}
\begin{aligned}
{\hat {\cal C}_{\alpha \beta }}(r,s) = \mathcal U(A,r,s)\left[ {{{{\rm{\bar \Psi }}}_\beta }(r - s/2),{{\rm{\Psi }}_\alpha }(r + s/2)} \right],
\end{aligned}
\end{equation}
where $r$ and $s$ are the center of mass and relative coordinates, respectively. To preserve the gauge invariance of the density operator presented in Eq. (\ref{DensityOperator}), the Wilson line factor is introduced. This factor is formally defined as follows:
\begin{equation}\label{Wilson}
	\begin{aligned}
\mathcal U(A,r,s) = \exp \left( {ies\int _{ - 1/2}^{1/2}{\rm{d}}\xi A(r + \xi s)} \right).
    \end{aligned}
\end{equation}
The covariant Wigner operator is defined as the Fourier transform of the density operator given in Eq.~\eqref{DensityOperator}
\begin{equation}\label{WignerOperator}
 \hat{\mathcal W}_{\alpha \beta} \left( r , p \right) = \frac{1}{2} \int d^4 s \
\mathrm{e}^{\mathrm{i} ps} \  \hat{\mathcal C}_{\alpha \beta} \left( r , s
\right).
\end{equation}
Considering the expected vacuum value from Eq.~\eqref{WignerOperator}, we obtain the covariant Wigner function $\mathbbm{W} \left( r,p \right)$
\begin{equation}\label{7}
	\begin{aligned}
\mathbbm{W}(r, p)  =\langle\Phi|\hat{\mathcal{W}}(r, p) |\Phi\rangle.
    \end{aligned}
\end{equation}
Following the Hartree-type approximation for the external field, and by employing the Dirac algebra of the Wigner function, we perform a spinor decomposition of the Wigner function in terms of the 16 covariant Wigner coefficients:
\begin{equation}
\mathbbm{W} = \frac{1}{4} \left( \mathbbm{1} \mathbbm{S} + \textrm{i} \gamma_5
\mathbbm{P} + \gamma^{\mu} \mathbbm{V}_{\mu} + \gamma^{\mu} \gamma_5
\mathbbm{A}_{\mu} + \sigma^{\mu \nu} \mathbbm{T}_{\mu \nu} \right). \,
\label{decomp}
\end{equation}
Here $\mathbb{S}$, $\mathbb{P}$, ${{\mathbb{V}}_{\mu }}$, ${{\mathbb{A}}_{\mu }}$ and ${{\mathbb{T}}_{\mu \nu }}$ are scalar, pseudoscalar, vector, axial-vector and tensor, respectively. In concert, $\mathbb{S}$ is related to the mass density, $\mathbb{P}$ is related to the condensate density, ${{\mathbb{V}}_{\mu }}$ is related to the net fermion current density, ${{\mathbb{A}}_{\mu }}$ is related to the polarization density, and ${{\mathbb{T}}_{\mu \nu }}$ is related to the electric dipole-moment density \citep{1991}.
According to Refs. \citep{Hebenstreit:2011wk,1991}, the dynamical equation for the Wigner function is
\begin{equation}\label{9}
	\begin{aligned}
{{D}_{t}}\mathbb{W}=-\frac{1}{2}{{\mathbf{D}}_{\mathbf{x}}}\left[ {{\gamma }^{0}}{\bm{\gamma}} ,\mathbb{W} \right]+im\left[ {{\gamma }^{0}},\mathbb{W} \right]-i\mathbf{P}\left\{ {{\gamma }^{0}}{\bm{\gamma}} ,\mathbb{W} \right\}.
    \end{aligned}
\end{equation}
Here, ${D_t}$, ${{\bf{D}}_{\bf{x}}}$, and ${\bf{P}}$ represent pseudo-differential operators.
To describe pair creation in a vacuum, we formulate it as an initial value problem. By considering the average energy of the Wigner function, we can derive the equal-time Wigner function
\begin{equation}
 \mathbbm{w} \left( \mathbf{x}, \mathbf{p}, t \right) = \int \frac{d p_0}{2 \pi}
\ \mathbbm{W} \left( r,p \right),
\end{equation}
where $\mathbf{x}$ and $\mathbf{p}$ denote the position and kinetic momentum of the particles, respectively. It should be noted that, for the studied problem of pair production, only the equal-time Wigner function can be employed to obtain the required evolution of the system. Consequently, all quantities discussed hereafter pertain to the components of the equal-time Wigner function.

Now we can apply the DHW formalism to the case of one spatial dimension add a time varying electric field $E(x,t)$ (1+1 dimensional scenario), in which
there are only four Wigner components, $\mathbbm{s}$, $\mathbbm{v}_{0}$, $\mathbbm{v}_{x}$ and $\mathbbm{p}$. Therefore, complete set of equations of motion for the Wigner components is reduced to four equations in this 1+1 dimensional case \citep{Hebenstreit:2011wk,Ababekri:2019dkl,1991}:
\begin{align}
 &D_t \mathbbm{s} - 2 p_x \mathbbm{p} = 0 , \label{pde:1}\\
 &D_t \mathbbm{v}_{0} + \partial _{x} \mathbbm{v} = 0 , \label{pde:2}\\
 &D_t \mathbbm{v} + \partial _{x} \mathbbm{v}_{0} = -2 m \mathbbm{p} , \label{pde:3}\\
 &D_t \mathbbm{p} + 2 p_x \mathbbm{s} = 2 m \mathbbm{v} , \label{pde:4}
\end{align}
where the pseudo-differential operator $D_t$ defined as
\begin{equation}\label{pseudoDiff}
 D_t = \partial_{t} + e \int_{-1/2}^{1/2} d \xi \,\,\, E_{x} \left( x + i \xi \partial_{p_{x}} \, , t \right) \partial_{p_{x}} .
\end{equation}
Note that, in our $1+1$ field configuration transverse momentum has little effect on the problem we are studying, then we ignore the transverse momentum of particle in for the simplicity of the theoretical and numerical calculation~\citep{Hebenstreit:2011wk,Malika:2021,Emin:2023,Amat:2023vwv,Bake:2024}.

Among the four Wigner components in the equations of motion given by Eqs. \eqref{pde:1}-\eqref{pde:4}, only two possess nonvanishing vacuum initial conditions \citep{Kohlfurst:2015zxi}:
\begin{equation}\label{vacuum-initial}
{\mathbbm s}_{vac} = - \frac{2m}{\Omega} \, ,
\quad  {\mathbbm v}_{vac} = - \frac{2{ p_x} }{\Omega} \,  ,
\end{equation}
where $\Omega=\sqrt{p_{x}^{2}+m^2}$ denotes the particle energy.
By explicitly subtracting these vacuum terms, the modified Wigner components can be written as follows:
\begin{equation}\label{vacuum-initial-}
{\mathbbm w}^{v} = {\mathbbm w} - {\mathbbm w}_{vac},
\end{equation}
where ${\mathbbm w}$ denotes the four Wigner components in our 1+1 case.

Finally, the particle number density in phase space can be formally defined as follows \citep{Hebenstreit:2011wk}:
\begin{equation}\label{PS}
n \left( x , p_{x} , t \right) = \frac{m  \mathbbm{s}^{v} \left( x , p_{x} , t \right) + p_{x}  \mathbbm{v}^{v} \left( x , p_{x} , t \right)}{\Omega \left( p_{x} \right)},
\end{equation}
where the total energy of the created particles is divided by the energy of an individual particle.
The position and  momentum distributions of the created particles can be obtained from $n \left( x , p_{x} , t \right)$ as follows:
\begin{equation}\label{PS1}
n \left( x , t \right) = \int d p_{x}~n \left( x , p_{x} , t \right),
\end{equation}
\begin{equation}\label{MS}
n \left( p_{x} , t \right) = \int \frac{d x}{2\pi}~n \left( x , p_{x} , t \right).
\end{equation}
The total particle yield was calculated by integrating over the entire phase space.
\begin{equation}\label{Num}
N\left(t \right) = \int \frac{d x}{2\pi} dp_x n \left( x , p_{x} , t \right).
\end{equation}
To obtain the nontrivial spatial dependence of the results on the field spatial scale $\lambda$ to facilitate comparison, the reduced quantities can be defined as follows:
\begin{equation}\label{ReNum}
\overline{n}\left(p_{x}, t \right) = \frac{n \left( p_{x} , t \right)}{\lambda} ~~~and~~~\overline{N}\left(t \right) = \frac{N\left(t \right)}{\lambda}.
\end{equation}

\subsection{External field model}\label{fields}

\begin{figure}[ht]\suppressfloats
\includegraphics[scale=0.4]{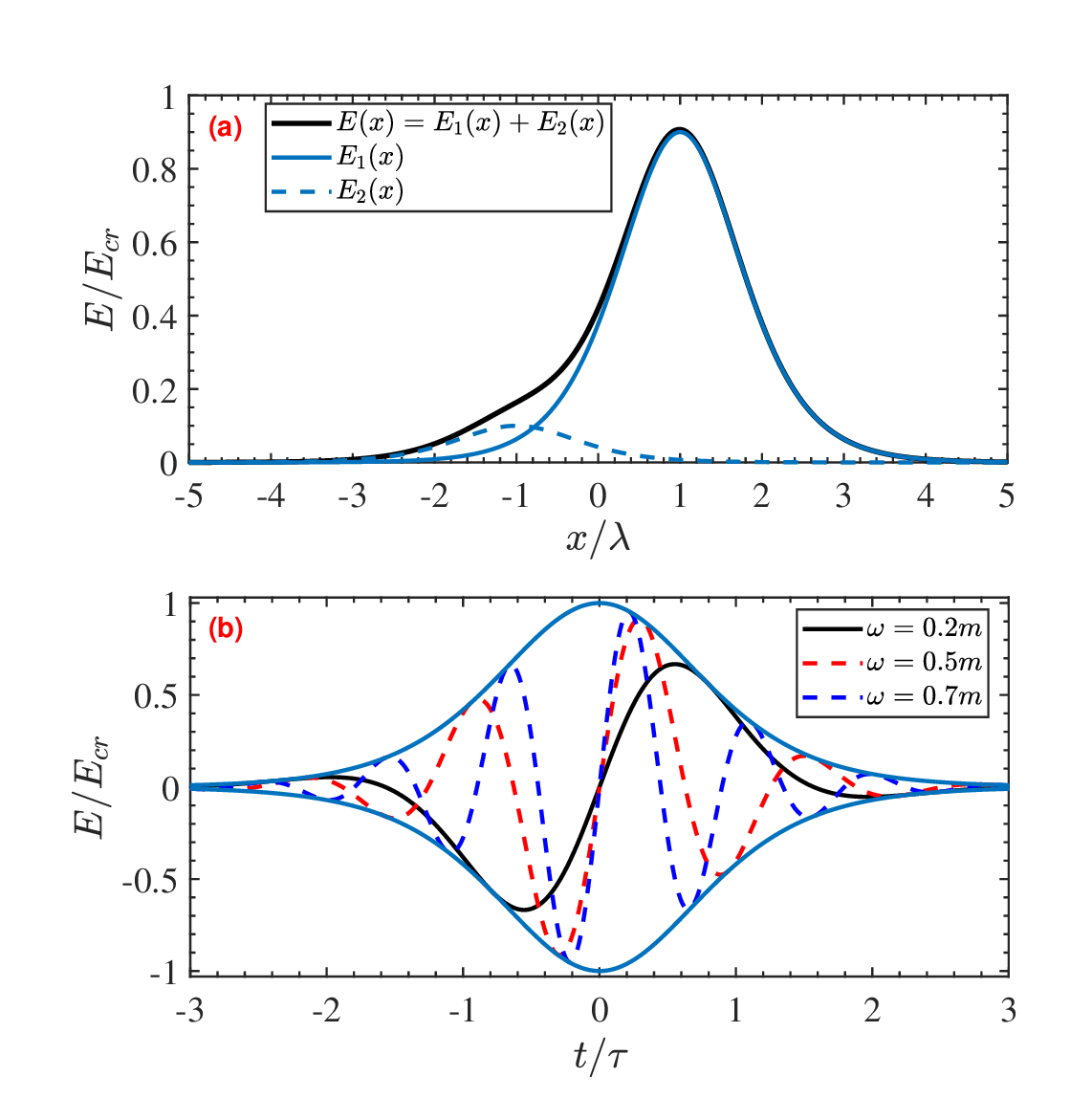}
\caption{The schematic diagram of the (a) spatial and (b) temporal profiles of the combined electric field, normalized to the critical field. Two symmetric, separated fields with fixed separation distance $d = 1m^{-1}$ is depicted by solid and dashed blue curves, whereas the black curve denotes the effective field in Fig.~\ref{fig1} (a). In Fig.~\ref{fig1} (b), the color-coded curves correspond to different original frequencies $\omega$. For the field parameters, $\epsilon_1=1.8$, $\epsilon_2=0.2$, $\epsilon=0.5$, the spatial scale $\lambda$ and frequency $\omega$ are modulated, while the temporal scale is maintained at a constant value of $\tau=10 m^{-1}$ during numerical computations.}
\label{fig1}
\end{figure}
Previous studies have demonstrated that the efficiency of pair creation depends on both the temporal and spatial characteristics of the external field, such as the temporal shape of the field~\citep{Oluk:2020,Chen:2024}, frequency chirp~\citep{Mamutjan:PLB,Malika:2021}. One of our prior investigations~\citep{Bake:2024} explored the influence of various spatial combinations of two spatially separated, space-dependent, temporally Sauter-type electric fields on the efficiency of pair production. The findings revealed that the spatial configuration of the electric field plays a critical role in enhancing the pair production rate and shaping the momentum spectrum. In this study, we analyze electron-positron pair production in 1+1 dimensions by considering a spatially asymmetric electric field with distinct temporal oscillations, as illustrated in Figs.~\ref{fig1} and \ref{fig5}. The field configuration is described as follows:
\begin{equation}\label{FieldMode}
\begin{aligned}
E\left(x,t\right)
&=\epsilon \, E_{cr} \left [ E_1(x)+E_2(x)\right] g(t)\\
&=\epsilon \, E_{cr} \left[ \epsilon_1 \mathrm{sech}^2 \left(\frac{x}{\lambda} - d\right) + \epsilon_2 \mathrm{sech}^2 \left(\frac{x}{\lambda} + d\right )\right]\\
&~~~~\times\mathrm{sech}^2 \left(\frac{t}{\tau}\right)\cos(\omega t),
\end{aligned}
\end{equation}
where $\lambda$ and $\tau$ represent the spatial and temporal scales of the field, respectively, and $\omega$ denotes the original frequency of the field. Note that, the equations of motion for the Wigner components given by Eqs.  \eqref{pde:1}--\eqref{pde:4} are numerically solvable only for a limited class of space-dependent electric fields. This is because the application of the pseudo-differential operator $D_t$ (Eq. (\ref{pseudoDiff})) introduces an error function that yields valid results only for specific field configurations, such as spatially homogeneous Gaussian or Sauter-type envelopes. Therefore, in this work, we constructed a spatially asymmetric effective field by combining two spatially separated electric fields with different intensities as given in Eq.~(\ref{FieldMode}). In Eq.~(\ref{FieldMode}), $E_{cr}$ represents the critical field strength; $E_1(x,t)$ and $E_2(x,t)$ denote the strengths of two spatially separated electric fields. In this study, based on the optimal values obtained from our previous research \citep{Bake:2024}, the strengths of the two fields are fixed at $\epsilon_1 = 1.8$ and $\epsilon_2 = 0.2$, respectively, with a spatial separation distance of $d = 1m^{-1}$.
In all the numerical calculations, we set $E_{0} = 0.5 E_{\text{cr}}$ ($\epsilon=0.5$) and adopted a temporal profile of $\mathrm{sech}^2(\frac{t}{\tau})$ (Sauter-type temporal envelope) with a fixed temporal scale of $\tau = 10 \, \text{m}^{-1}$. It is worth noting that the field was oriented along the $x$-axis, while its strength varied as a function of both $x$ and $t$.
In Fig. \ref{fig1} (a), it is evident that the superposition of two spatially separated fields generates a new asymmetric effective field, which will substantially alter the dynamics of pair production and the particle phase-space distribution. In Fig. \ref{fig1} (b), the temporal component of the field is illustrated for various frequencies.
By employing this idealized field model, we will examine the effects of a spatially asymmetric electric field with a specific time-dependent oscillation on the created particle number density, momentum distribution, and the total reduced particle number in phase space.
\begin{figure}[ht]\suppressfloats
\includegraphics[scale=0.45]{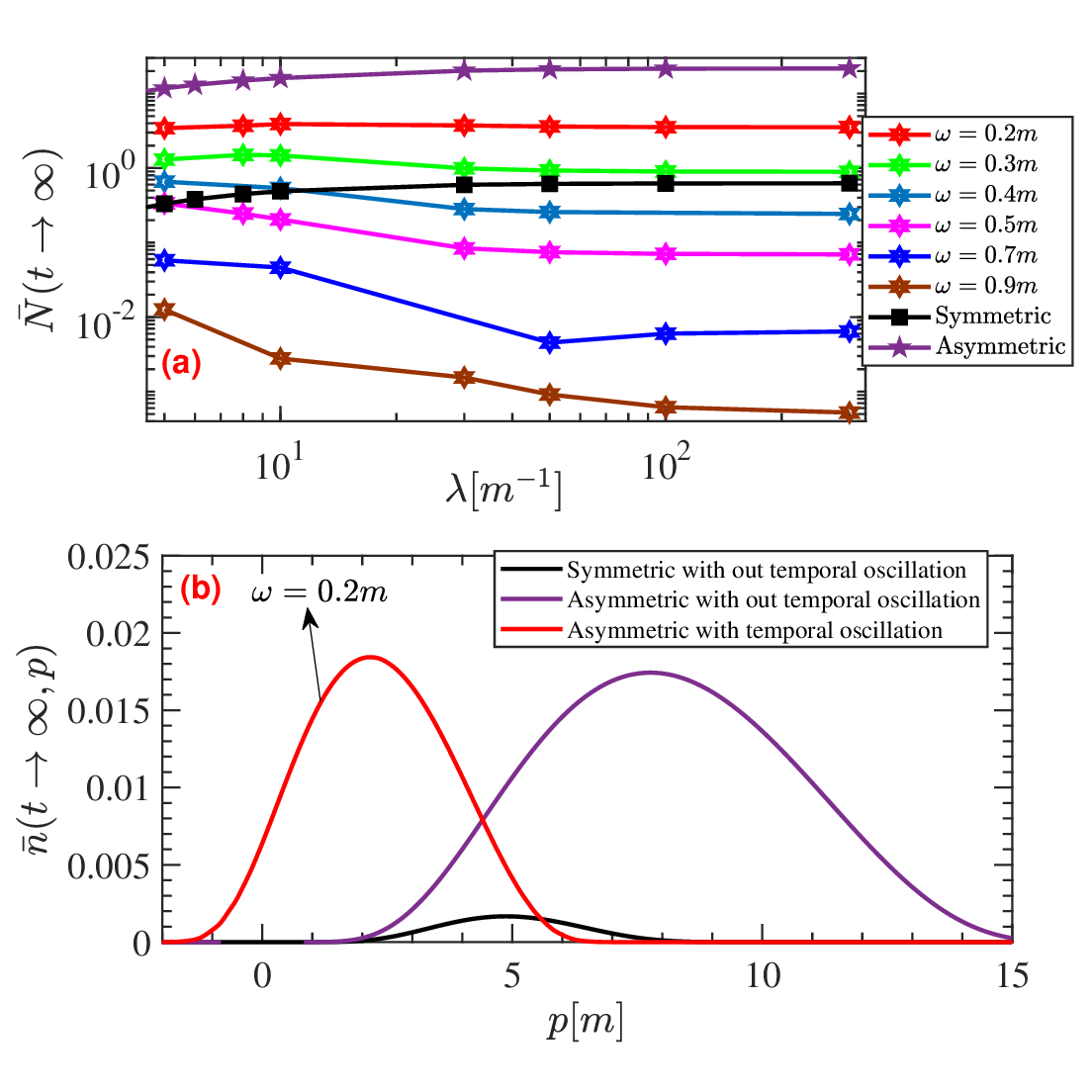}
\caption{(a) Reduced total particle number $\bar{N}$ at various spatial extents $\lambda$ for electric fields with different central frequencies in Eq. (\ref{FieldMode}) and with various spatial configurations. (b) Momentum spectrum of particles for various fields with fixed spatial scale $\lambda=50m^{-1}$. The other parameters are the same as in the Fig. \ref{fig1}.}
\label{fig2}
\end{figure}

\section{Results and analysis}\label{results}
\begin{figure*}[ht]\suppressfloats
\includegraphics[scale=0.55]{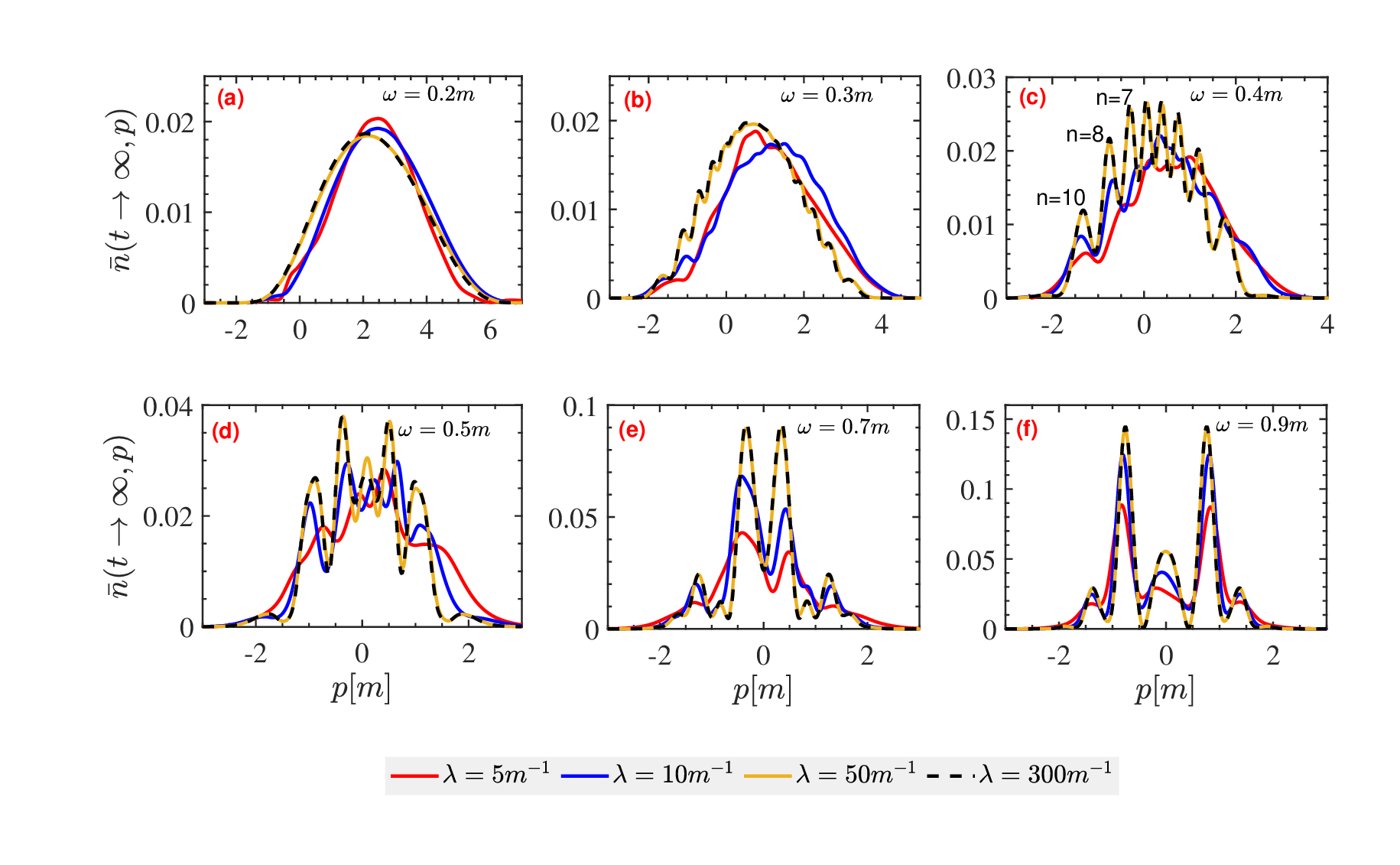}
\caption{The momentum distribution at typical four spatial extents $\lambda$ for a spatially asymmetric field with various temporal frequencies. The other parameters are the same as in the Fig. \ref{fig1}.}
\label{fig3}
\end{figure*}
In this section, we present the results calculated using the DHW formalism described in Section \ref{DHWformalism}, based on the external field configurations outlined in Section \ref{fields}. Additionally, we analyze these results in detail for various field parameters.

We first compare the reduced total particle number $\overline{N}$ in different field configurations, as shown in Fig. \ref{fig2} (a). The results indicate that $\overline{N}$ is higher for the spatially asymmetric field without temporal oscillation (purple line) compared to the spatially symmetric field without temporal oscillation (black line) and the spatially asymmetric field with temporal oscillation (red line). This suggests that optimal pair production can be achieved using spatially asymmetric electric fields without temporal oscillation.
However, in realistic field configurations, the field must be considered spatiotemporally inhomogeneous and oscillates with a specific frequency over time, and these inhomogeneities can significantly impact the vacuum pair production precess.
Therefore, it is of critical importance to examine the effects of spatially asymmetric fields with varying temporal oscillation frequencies on pair production. The reduced total particle number $\overline{N}$ decreases as the field frequency increases, as shown in Fig. \ref{fig2} (a). This result is not entirely consistent with the intuition that higher field frequencies correspond to higher photon energy, which in turn leads to a greater number of pairs.
This is because, when fast electric fields (high-frequency fields) that are spatially inhomogeneous are superimposed, the dynamical assistance is diminished due to the increased momentum transfer from the fast fields, which widens the vacuum energy gap \citep{Aleksandrov2018,Torgrimsson2018}. Consequently, dynamical assistance is essentially maximized when perturbations are purely time-dependent or space-dependent. Figure \ref{fig2} (b) plots the particle momentum spectrum corresponding to various field configurations with the same spatial scale $\lambda=50~m^{-1}$. We observed that the distribution of the omentum spectrum exhibits significant sensitivity to variations in field configurations. The peak value of the number density is higher under an asymmetric temporally oscillating field, whereas it broadens and shifts toward the positive momentum direction in cases of purely spatial dependence.

As is well known, with the increase in the frequency of electric fields, if it reaches a significant level, it can enable perturbative pair creation involving a finite number of photons. The qualitative characteristics of pair creation undergo substantial changes depending on whether nonperturbative or perturbative mechanisms are dominant.
We investigate the momentum spectra of the produced pairs for a spatially asymmetric electric field with varying spatial scales and temporal oscillation frequencies, as illustrated in Fig. \ref{fig3}. For a small frequency of $\omega = 0.2m$, the pairs are predominantly created via a nonperturbative tunneling process due to the relatively smaller Keldysh parameter. Consequently, the momentum spectrum of the produced pairs exhibits a soft Gaussian distribution with approximate bilateral symmetry, and the result is similar across all spatial scales of the field, as illustrated in Fig. \ref{fig3} (a). As the field frequency increases to $\omega=0.3m$, for smaller spatial scales of $\lambda=5m^{-1}$ and $\lambda=10m^{-1}$, the shape of the momentum spectra remains identical to that shown in Fig. \ref{fig3} (b). In contrast, for larger spatial scales ($\lambda=50 m^{-1}$ and $\lambda=300 m^{-1}$), the momentum spectrum begins to exhibit a multi-peak distribution, as illustrated in Fig. \ref{fig3} (b). This is a prominent characteristic of the perturbative multiphoton pair production process.
With a further increase in the field frequency, the peak value of the momentum spectrum increases, the momentum spectrum narrows and shifts toward the negative $p$ direction.

In the higher-frequency cases, as illustrated in Figs. \ref{fig3} (c)-(f), due to the perturbative pair creation being only weakly suppressed by a power law, the momentum spectrum acquires a harder oscillation component. The high-frequency components of the electric field, which are more closely associated with the perturbative mechanism, enhance pair creation compared to the naive expectation based on the non-perturbative mechanism by absorb more high energy photons.
One can calculate the absorbed photon number $n$ by employing the energy conservation equation $n\omega=2\sqrt{m^2_*+p^2}$ with the effective mass $m_*=m[1+(mE_0/\omega E_{cr})^2/2]^{1/2}$ \citep{CK2015}. For instance, the three peaks in Fig. \ref{fig3} (c) correspond to seven, eight, and ten photon pair productions.
The pronounced oscillations observed in the momentum spectrum at higher frequencies, particularly for $\omega=0.4m$ and $\omega=0.5m$, can be attributed to the quantum interference effect. Specifically, pair creation does not occur at a single instance but rather at multiple times. Electrons created at different times exhibit a non-vanishing relative phase. Consequently, pair creation events at distinct times interfere with one another, leading to constructive or destructive interference in the pair creation process. This interference modifies the momentum spectra, resulting in characteristic oscillatory patterns.

For further increases in frequency to $\omega=0.7m$ and $\omega=0.9m$, the interference effect is likely to be suppressed, and peak splitting occurs, as shown in Figs. \ref{fig3} (e) and (h). When fast electric fields (high frequency fields) that are superimposed exhibit spatial inhomogeneity, the pair production effect is reduced because the momentum provided by the fast fields widens the gap, as discussed in Fig. \ref{fig2} (a). On the other hand, quantum-interference effects also tend to be suppressed by spatial inhomogeneities \citep{Liliejuan:2021,Malika:2021}. Indeed, it becomes challenging for pairs created at different spacetime points to occupy the same phase-space after undergoing inhomogeneous fast acceleration. Consequently, different pairs rarely interfere with each other, leading to the suppression of oscillating structures in momentum spectra \citep{Ababekri:2019dkl}.
Meanwhile, following their creation, the generated pairs are accelerated by the electric field. If the spatial extent of the electric field is sufficiently large, the pairs remain within the field after creation and consequently acquire substantial momentum from it. Thus, spatial inhomogeneities, by altering the dynamics post-pair creation, influence the momentum spectra, resulting in self-bunching \citep{Hebenstreit:2011wk,Ababekri:2019dkl}. Consequently, we observe pronounced symmetric splitting of the momentum spectrum for large spatial scales $\lambda=50m^{-1}$ and $\lambda=300m^{-1}$, as depicted in Figs. \ref{fig3} (e) and (h), which corresponds to the particle self-bunching.

\begin{figure}[ht]\suppressfloats
\includegraphics[scale=0.45]{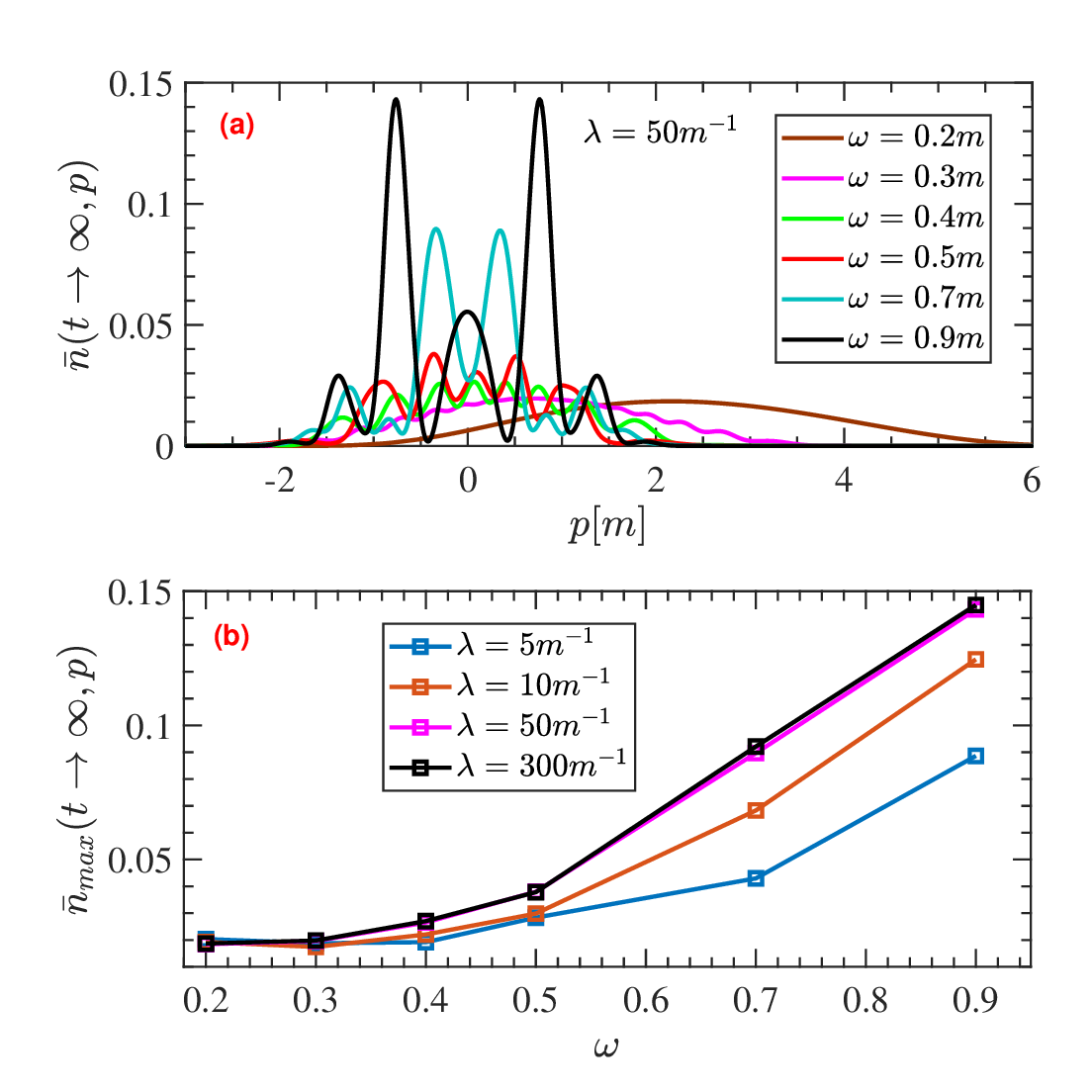}
\caption{(a) The momentum distribution for various temporal frequencies $\omega$ in a spatially asymmetric field with a fixed spatial scale of $\lambda = 50m^{-1}$. (b) The maximum reduced particle density $\overline{n}_{max}$ as a function of temporal frequency $\omega$ for four typical spatial extents $\lambda$. The other parameters are same as in Fig. \ref{fig1}.}
\label{fig4}
\end{figure}

In Fig. \ref{fig4} (a), the frequency dependence of the momentum spectrum is illustrated for a fixed, relatively large spatial scale $\lambda=50m^{-1}$ (indicating a high degree of spatial asymmetry in the field). It is observed that the momentum spectrum broadens and exhibits nonperturbative characteristics at low frequencies ($\omega=0.2m$, $0.3m$). Conversely, at higher frequencies ($\omega=0.4m$, $0.5m$), the spectrum narrows and demonstrates distinct perturbative oscillation characteristics. Additionally, it is centered around $p=0$ and exhibits a self-bunching form with a higher peak value for high frequency of $\omega=0.7m$, $0.9m$.
As the frequency $\omega$ increases, the number of cycles in the envelope of our field configuration also increases, as illustrated in Fig. \ref{fig1} (b). This alteration can modify the timing of pair creation events, thereby influencing the interference pattern and eventually leading to a complex pattern of the momentum spectrum. Such behavior is notably distinct from that of the applied electric field. These findings suggest that by carefully designing the characteristics of the fields such as their spatial and temporal profiles, as well as the cycle structure of a laser pulses it may be possible to control both the timing of pair production and the momentum spectrum pattern. This is crucial for optimizing pulse shaping strategies aimed at maximizing the Schwinger effect in future experimental setups.

\begin{figure*}[ht]\suppressfloats
\includegraphics[scale=0.5]{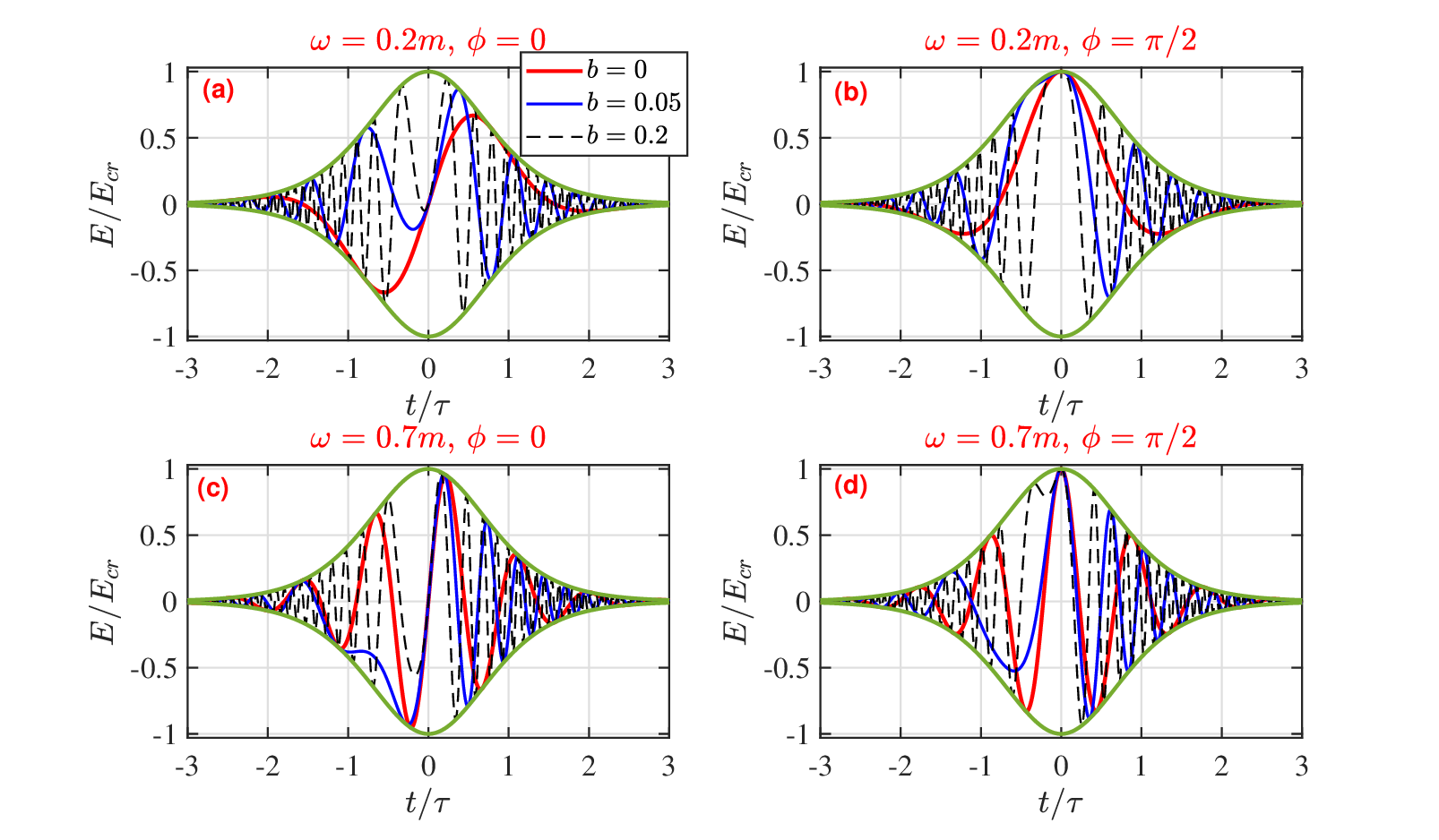}
\caption{The schematic diagram illustrates the temporal component of the electric field for different frequency chirp parameters $b$, featuring an asymmetric spatial profile as shown in Fig. \ref{fig1} (a). The first row corresponds to the low-frequency case ($\omega=0.2m$), while the second row represents the high-frequency case ($\omega=0.7m$). The first column depicts the scenario with an initial phase of $\phi=0$, and the second column shows the scenario with an initial phase of $\phi=\pi/2$. The other parameters are same as in Fig. \ref{fig1}.}
\label{fig5}
\end{figure*}

From Fig. \ref{fig2} (a), Fig. \ref{fig3}, and Fig. \ref{fig4} (b), we can systematically analyze the influence of the spatial extent $\lambda$ of the electric field on vacuum pair production. It is observed that for large values of $\lambda$ ($\lambda \geq 50m^{-1}$), the reduced total particle number $\overline{N}$ remains nearly constant across various field configurations and frequencies, as depicted in Fig. \ref{fig2} (a). Furthermore, the maximum particle density exhibits a clear increase with both $\omega$ and $\lambda$, with identical trends observed for $\lambda = 50m^{-1}$ and $\lambda = 300m^{-1}$, as illustrated in Fig. \ref{fig4} (a). These phenomena can be attributed to the quasi-homogeneous nature of the electric field configuration at large $\lambda$.
The quasi-homogeneous results were analyzed using the local density approximation as discussed in Refs. \citep{Hebenstreit:2011wk,Ababekri:2019dkl,Mamutjan:PLB}. For large spatial scales, the homogeneous field approximation can be employed, and the momentum distribution (or reduced particle density $\bar{n}$) can be computed by summing the results for homogeneous fields with different field strengths \citep{Ababekri:2019dkl}: $\tilde{n}(p, t \rightarrow \infty) = \sum_{x} \frac{n(\epsilon(x)| p, t \rightarrow \infty)}{\lambda}$, where $n(\epsilon(x)| p, t \rightarrow \infty)$ represents the momentum distribution of a homogeneous field $E(t)$ with an effective field strength given by $E(t) = \epsilon(x) E_{cr} \mathrm{sech}^2 \left(\frac{t}{\tau}\right)$.
It is important to note that the homogeneous field approximation used in this study is only valid when the pair-formation length $l=2m/|e \epsilon E_{cr}|$ is significantly smaller than the spatial width $\lambda$ of the external field \citep{Hebenstreit:2011wk}. External fields with a large spatial scale $\lambda$ can readily satisfy this condition. Figure \ref{fig3} demonstrates the quasi-homogeneous characteristics of the field for cases with a large spatial extent (large $\lambda$). For $\lambda=50m^{-1}$ and $\lambda=300m^{-1}$, the momentum spectrum exhibits an identical distribution. This indicates that, under the conditions of our study, the uniform field approximation holds for $\lambda \geq 50m^{-1}$, allowing the electric field to be treated as quasi-homogeneous. Consequently, the momentum spectrum remains unchanged for larger spatial extents. Therefore, we have not considered even larger spatial scales in this work.

Many previous studies have investigated the significant impact of frequency chirp on pair production. Specifically, the introduction of a chirp causes the original field frequency to vary with time in certain modes. Figure \ref{fig5} illustrates the temporal evolution of the electric field under various chirp conditions. It is evident that different chirp parameters alter the oscillation patterns and cycle numbers within the same temporal envelope. This phenomenon affects the pair production process and subsequently modifies post-particle dynamics, changing the momentum spectrum.
However, most of these previous studies have focused on spatially homogeneous or spatially symmetric inhomogeneous fields. Consequently, the influence of spatially asymmetric fields with temporal frequency chirp on pair production remains unclear. Therefore, in this work, we introduce a linear frequency chirp to a spatially asymmetric field. An effective frequency for the linearly chirped electric field is defined, which can be expressed by the following equation: $\omega_{eff} = \omega + bt$. Subsequently, the temporal component of the external field presented in Eq. (\ref{FieldMode}) is formulated as $g(t) = \mathrm{sech}^2 \left(\frac{t}{\tau}\right)\cos(\omega t + bt^2)$, where $b$ represents the chirp parameter.

\begin{figure*}[ht]\suppressfloats
\includegraphics[scale=0.55]{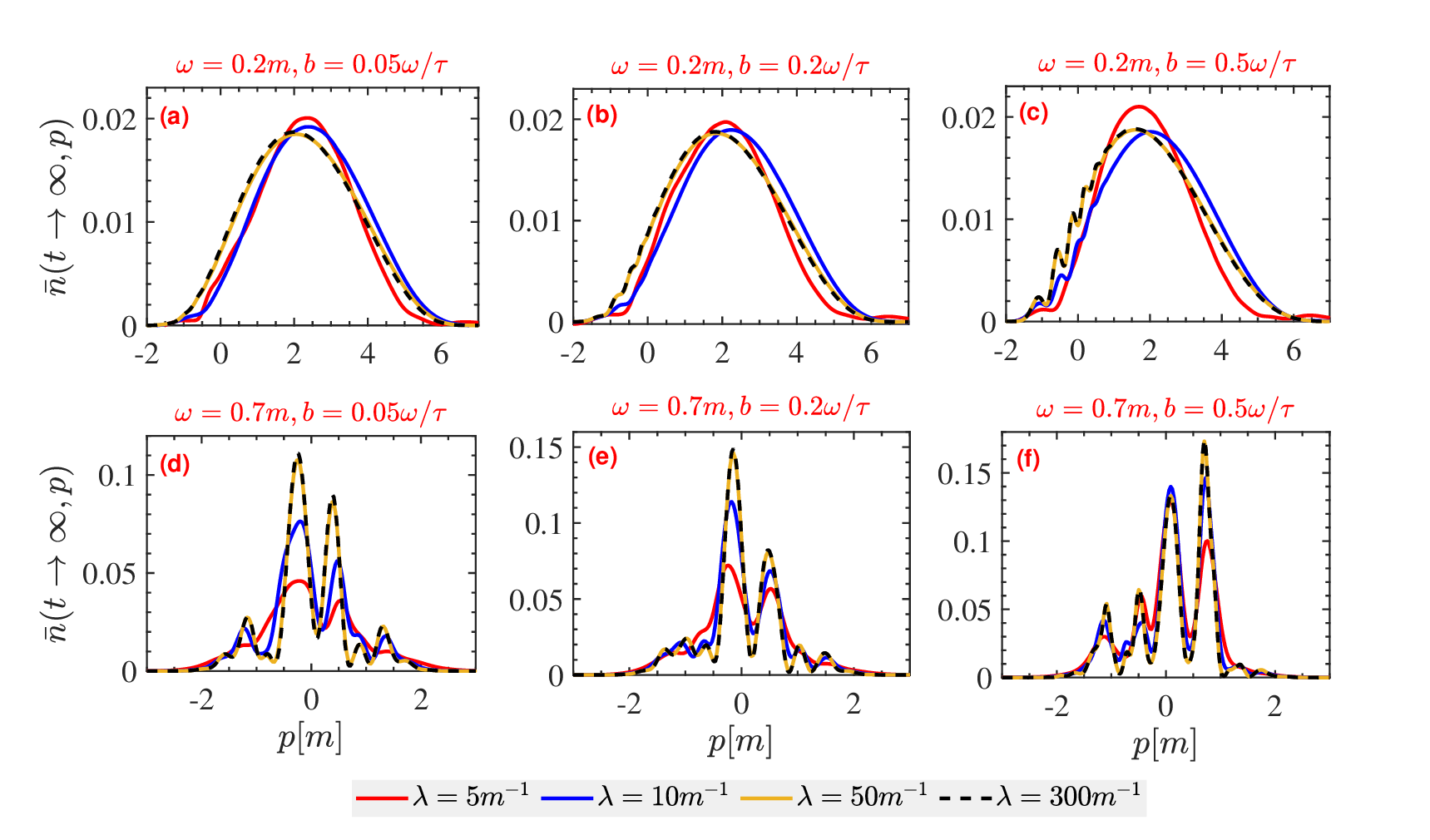}
\caption{The momentum distribution at typical four spatial extents $\lambda$ for a spatially asymmetric field for low (first row) and high (second row) temporal frequencies with various chirp parameter $b$. The other parameters are the same as in the Fig. \ref{fig1}.}
\label{fig6}
\end{figure*}

The momentum spectra of the created particles for low-frequency ($\omega=0.2m$) and high-frequency ($\omega=0.7m$) chirped fields with various chirp parameters $b$ and spatial scales are presented in Fig. \ref{fig6}. It was found that the patterns and peak values of the momentum spectra for different chirp parameters exhibit distinct differences. For the low-frequency chirped field with chirp parameters $b=0.05\omega/\tau, 0.2\omega/\tau$, the momentum spectra do not show significant differences compared to the case with vanishing chirp ($b=0$) (see Fig. \ref{fig3} (a)), as illustrated in Figs. \ref{fig6} (a) and (b).
As the chirp parameter is increased to $b=0.5\omega/\tau$, the momentum spectrum begins to exhibit nonperturbative multi-photon oscillations, with a slight increase in peak value as shown in Fig. \ref{fig6} (c). This occurs because, although the original field frequency is relatively small ($\omega=0.2m$), the introduction of the chirp modifies the effective frequency $\omega_{eff}$, which in this case increases with the chirp parameter $b$. Consequently, oscillations in the momentum spectrum can be observed at larger spatial scales when the chirp parameter is increased. Such oscillations are typically observed in an unchirped electric field only for a higher original frequency, such as $\omega=0.3m$ (see Fig. \ref{fig3} (b)).

The momentum spectrum for a higher frequency ($\omega=0.7m$) electric field with varying chirp parameter $b$ is presented in Figs. \ref{fig6} (d)-(f). It is evident that both the peak value and the oscillatory behavior of the momentum spectrum are enhanced as the chirp parameter $b$ increases. The oscillatory pattern in the momentum spectrum can be attributed to the interference between two temporally separated pair creation events, compounded by the complex oscillations induced by a large chirp parameter. Electrons created at different times exhibit a non-vanishing relative phase, which leads to interference during the pair creation process. This relative phase is momentum-dependent, with momentum being modulated by the electric field. As a result, the interference alternates between constructive and destructive states, thereby inducing characteristic variations in the momentum spectra that are significantly influenced by the form of the applied electric field.
It is also observed from Figs. \ref{fig6} (d)-(f) that, upon introducing the chirp, the peak value of the momentum spectrum is enhanced while the symmetry of the momentum spectrum is broken. In the absence of chirping, for large original frequencies and spatial scales, the momentum spectrum remains symmetric about $p=0$, as illustrated in Figs. \ref{fig3} (d)-(f). However, in the presence of a chirped field, the symmetry is disrupted, and the highest peak shifts toward the positive momentum side, as shown in Figs. \ref{fig6} (d)-(f). This phenomenon may be attributed to the post-dynamics of created pairs influenced by the chirp effect of the electric field, which leads to asymmetric temporal oscillations of the field within the envelope (see Fig. \ref{fig5}). The broken symmetry under the chirped field condition results in distinct particle dynamics following creation. These findings indicate that the peak of the momentum spectrum shifts progressively toward the positive $p$ region with an increase in the chirp parameter $b$.

\begin{figure}[ht]\suppressfloats
\includegraphics[scale=0.45]{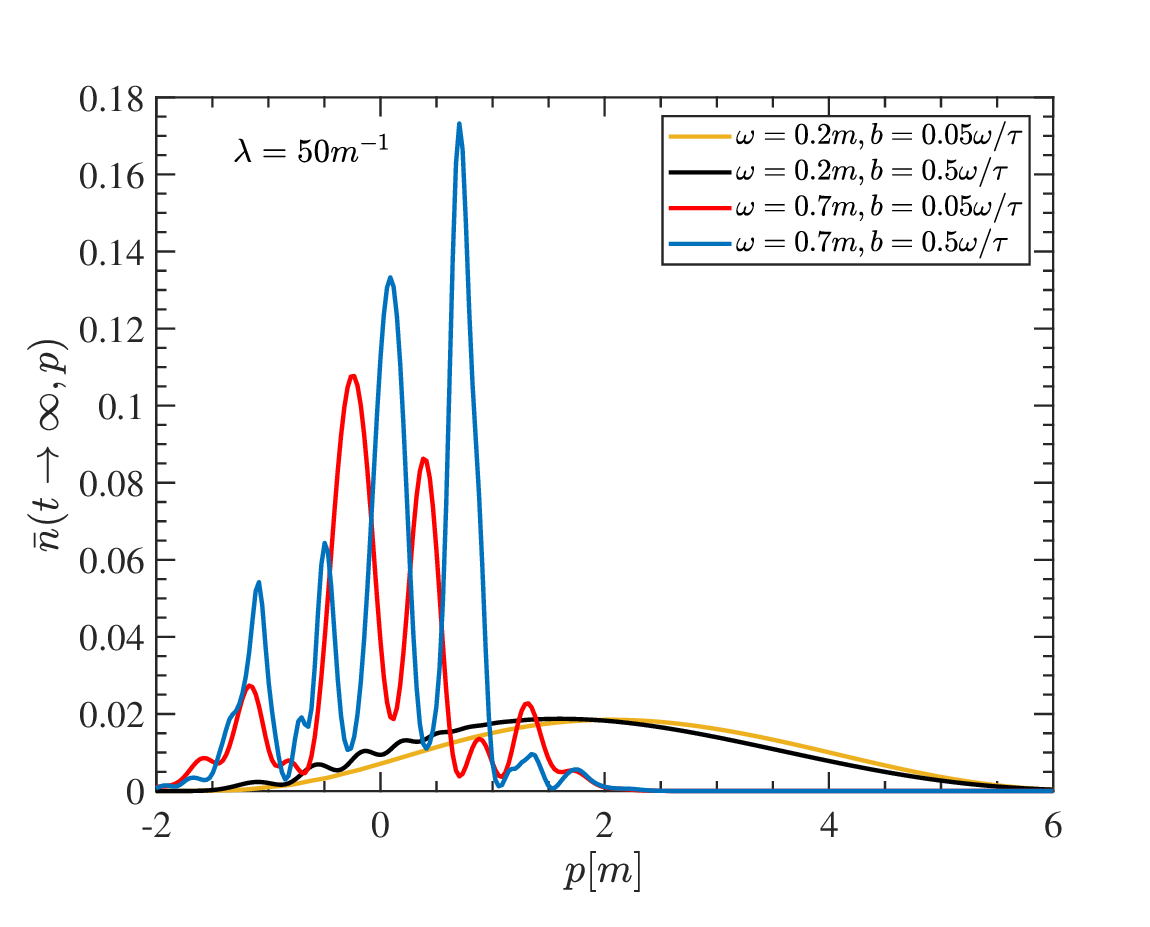}
\caption{The momentum distribution for two different temporal frequencies $\omega$ with various chirp parameter $b$ and a fixed spatial scale of $\lambda = 50m^{-1}$. The other parameters are the same as in the Fig. \ref{fig1}.}
\label{fig7}
\end{figure}

As shown in Fig. \ref{fig3}, for large spatial scales ($\lambda \geq 50m^{-1}$), the spatial inhomogeneous approximation can be applied. Consequently, in Fig. \ref{fig7}, we present the momentum spectrum under a fixed spatial scale of $\lambda = 50m^{-1}$, considering low and high-frequency electric fields with distinct chirp parameters $b=0.05\omega/\tau$ (small chirp) and $b=0.5\omega/\tau$ (large chirp). It was observed that the peak value of number density increases with both the original frequency and the chirp parameter. Specifically, for the high-frequency case with a large chirp parameter ($\omega=0.7m,~b=0.5\omega/\tau$), the number density is approximately nine times higher than that of the low-frequency case with the same chirp parameter ($\omega=0.2m,~b=0.5\omega/\tau$). Additionally, for the low-frequency scenario, the momentum spectrum is broader and exhibits weak oscillations only when the chirp parameter is large ($\omega=0.2m,~b=0.5\omega/\tau$). In contrast, for the high-frequency case, the momentum spectrum demonstrates strong oscillations regardless of whether the chirp parameter is small or large. This oscillatory pattern in the momentum spectrum arises from interference effects, as previously discussed.
\begin{figure}[ht]\suppressfloats
\includegraphics[scale=0.45]{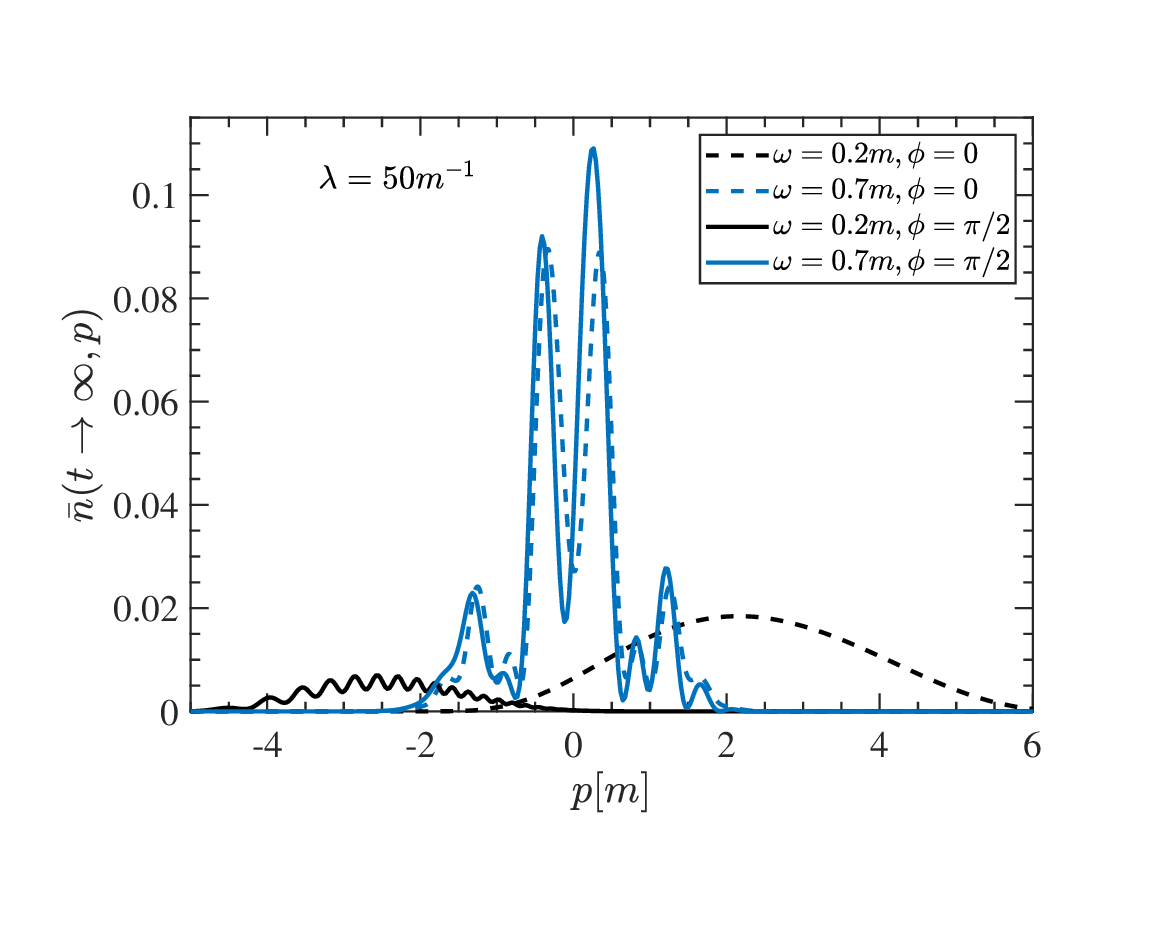}
\caption{The momentum distribution for two different initial phase of  $\phi=0$ and $\phi=\pi/2$ with a fixed spatial scale of $\lambda = 50m^{-1}$. The other parameters are the same as in the Fig. \ref{fig1}.}
\label{fig8}
\end{figure}

Finally, we investigate the effect of the initial phase of spatially asymmetric electric fields on the efficiency of pair production. As shown in Fig. \ref{fig5}, when the initial phase changes from $\phi=0$ to $\phi=\pi/2$, the position of the maximum electric field shifts to $t = 0$. This shift alters the effective strength of the field and the number of cycles within a fixed envelope, thereby influencing the efficiency of pair production and the distribution of the momentum spectrum. The results for unchirped electric fields, at both low and high frequencies with two selected initial phases, are presented in Fig. \ref{fig8}.
We systematically compared and analyzed the impact of different initial phase values on the momentum distribution by assigning two specific initial phase values, namely $\phi=0$ and $\phi=\pi/2$. Our findings indicate that, for the low-frequency case ($\omega=0.2m$), the peak of the momentum distribution is located in the negative momentum region and shifts to the positive momentum region as the initial phase value transitions from $\phi=0$ to $\phi=\pi/2$. Additionally, the momentum spectrum corresponding to $\phi=0$ exhibits an oscillatory pattern. In contrast, for the high-frequency case ($\omega=0.7m$), the momentum spectra display a similar distribution but with a significantly larger peak value when the initial phase is set to $\phi=\pi/2$. This suggests that the efficiency of pair production increases with the increasing initial phase.
This is attributed to the higher field strength in the $\phi=\pi/2$ case. However, the result is contrary for the low-frequency scenario, where the peak value of the $\phi=0$ case is larger than that of the $\phi=\pi/2$ case. This may be due to the dominance of different pair production mechanisms at various phases, which arises from the change in the cycle number. For instance, in the low-frequency case with $\phi=\pi/2$, there are strong oscillations in the momentum spectrum as shown in Fig. \ref{fig8}, which results from the multiphoton perturbation process. Conversely, for the $\phi=0$ case, the momentum spectrum of the produced pairs exhibits a soft Gaussian distribution with approximate bilateral symmetry, which originates from the typical nonperturbative tunneling process.

\section{Conclusion}\label{summary}
In conclusion, we have systematically studied vacuum pair production in spatially asymmetric electric fields with time-dependent oscillations using the Dirac-Heisenberg-Wigner (DHW) formalism. Our study highlights that spatially asymmetric, time-oscillating electric fields significantly influence vacuum pair production. The findings suggest that carefully designed field configurations can enhance pair production efficiency and control momentum spectra. These results provide valuable insights for future high-intensity laser experiments and further studies on vacuum pair production. Key findings include:
\begin{itemize}
    \item \textbf{Spatial asymmetry} enhances pair production rates compared to symmetric fields, with the highest yields achieved for large spatial scales ($\lambda \geq 50\,m^{-1}$) where the field approaches quasi-homogeneity.
    \item \textbf{Temporal parameters} critically influence the production mechanism: Low-frequency fields ($\omega \lesssim 0.3\,m$) favor tunneling, while high frequencies ($\omega \gtrsim 0.4\,m$) trigger multiphoton processes, evidenced by oscillatory momentum spectra.
    \item \textbf{Frequency chirps} ($b$) break spectral symmetry and amplify yields, with a ninefold increase observed for $\omega = 0.7\,m$ and $b = 0.5\omega/\tau$.
    \item \textbf{Initial phases} modulate interference effects, with $\phi = \pi/2$ maximizing yields in high-frequency regimes.
\end{itemize}
These results underscore the importance of spatiotemporal field engineering for experimental realizations of the Schwinger effect. Future work should explore multidimensional field configurations and quantum interference dynamics to further optimize pair production efficiency.

\begin{acknowledgments}
This work was supported by the National Natural Science Foundation of China (NSFC) (Grant No. 12265024), and the Special Training Program of Science and Technology Department of Xinjiang China (Grant No. 2024D03007).
O. Olugh was financially supported by the Natural Science Foundation of the Xinjiang Uyghur Autonomous Region, PR. China under grant number 2024D01A54.	
\end{acknowledgments}

\end{document}